\documentclass[twocolumn,showpacs,preprintnumbers,amsmath,amssymb]{revtex4-1}
\usepackage{comment}
\usepackage{graphicx,graphics,wrapfig,rotating}         
\usepackage{epsfig,subfigure}
\usepackage{dcolumn}                                    
\usepackage{bm,fancybox}
\usepackage{amsmath,amssymb}                          
\usepackage{times,euscript,oldgerm}              %
\usepackage[english]{babel}                      %
\usepackage{psfrag}
\usepackage{hyperref}

\begin{document}

\title{Onset of a Quantum Phase Transition with a Trapped Ion Quantum Simulator}
\author
{R. Islam{$^{1}$}, E. E. Edwards{$^{1}$}, K. Kim{$^{1}$}, S. Korenblit{$^{1}$}, C. Noh{$^{2}$}, H. Carmichael{$^{2}$}, G.-D.Lin{$^{3}$}, L.-M. Duan{$^{3}$}, 
C.-C. Joseph Wang{$^{4}$}, J. K. Freericks{$^{4}$}, C. Monroe{$^{1}$}}
\affiliation{$^1$ Joint Quantum Institute, University of Maryland Department of
Physics and \\
                    National Institute of Standards and Technology, College
Park, MD  20742 \\
             $^2$ Department of Physics, University of Auckland, Private Bag 92019, Auckland, NZ\\
             $^3$ MCTP and Department of Physics, University of Michigan, Ann
Arbor, Michigan 48109\\
             $^4$ Department of Physics, Georgetown University , Washington, DC
20057 }
\date{\today}
 \begin{abstract}
A quantum simulator is a well controlled quantum system that can simulate the behavior of another quantum system which may require exponentially large classical computing resources to understand otherwise.  In the 1980s, Feynman proposed the use of quantum logic gates on a standard controllable quantum system to efficiently simulate the behavior of a model Hamiltonian \cite{Feynman,Lloyd96}. Recent experiments using trapped ions \cite{Schaetz08,KimPRL,KimNature,EdwardsPRB} and neutral atoms  \cite{Greiner_arXiv} have realized quantum simulation of Ising model in presence of external magnetic fields, and showed almost arbitrary control in generating non-trivial Ising coupling patterns \cite{KimPRL}. Here we use laser-cooled trapped $^{171}\rm{Yb}^{+}$ ions to simulate the emergence of magnetism in a system of interacting spins  \cite{Porras04, DengPorras05, TaylorPRA} by implementing a fully-connected non-uniform ferromagnetic Ising model in a transverse magnetic field.  To link this quantum simulation to condensed matter physics, we measure scalable correlation functions and order parameters appropriate for the description of larger systems, such as various moments of the magnetization. By increasing the Ising coupling strengths compared with the external field, the crossover from paramagnetism to ferromagnetic order sharpens as the system is scaled up from $N = 2$ to $9$ trapped ion spins.  This points toward the onset of a quantum phase transition that should become infinitely sharp as the system approaches the macroscopic scale.  We compare the measured ground state order to theory, which may become intractable for non-uniform Ising couplings as the number of spins grows beyond 20- 30 \cite{Lanczos} and even NP complete for a fully-connected frustrated Ising model \cite{Barahona}, making this experiment an important benchmark for large-scale quantum simulation.
\end{abstract}

\maketitle

We find the ground state of the following Hamiltonian for interacting spin$-1/2$ systems :
\footnote{In the experiment, we actually follow the highest excited state of the Hamiltonian $-H$ \cite{Schaetz08,KimPRL}, which is formally equivalent to the ground state of Eq. \ref{Ham}.} 
\begin{equation}
\label{Ham}
H=-\frac{1}{N}\sum_{i<j}J_{i,j}\sigma_{x}^{i}\sigma_{x}^{j}-B\sum_{i}\sigma_{y}^{i}
\end{equation}
where $\sigma_{\alpha}^{i}$ is the Pauli matrix for the $i^{\rm{th}}$ spin ($\alpha=x,y,z$ and $i=1,2,...,N$) and $J_{i,j}>0$ is the ferromagnetic (FM) Ising coupling matrix, with $J = \langle J_{i,j} \rangle$.  Our experiment is performed according to adiabatic quantum simulation protocol \cite{Farhi01} where the dimensionless coupling  $B/|J|$ is tuned slowly enough so that the system follows instantaneous eigenstates of the changing Hamiltonian \cite{KimNature, EdwardsPRB}. As $B/|J|\to\infty$, the ground state has all spins polarized along the magnetic field, or paramagnetic along the Ising direction $X$. In the other limit $B/|J|=0$, the spins order according to the Ising couplings and the ground state is a superposition of FM states $|\uparrow\uparrow...\uparrow\rangle$ and $|\downarrow\downarrow...\downarrow\rangle$ where $|\uparrow\rangle$ and $|\downarrow\rangle$ are eigenstates of $\sigma_{x}$.
We characterize the magnetic order in the system by measuring various correlation functions between all $N$ spins, including the probability of FM occupation and the second and fourth moments of the total magnetization.

\begin{figure*}
\begin{flushleft}
 
\includegraphics[width=1\linewidth]{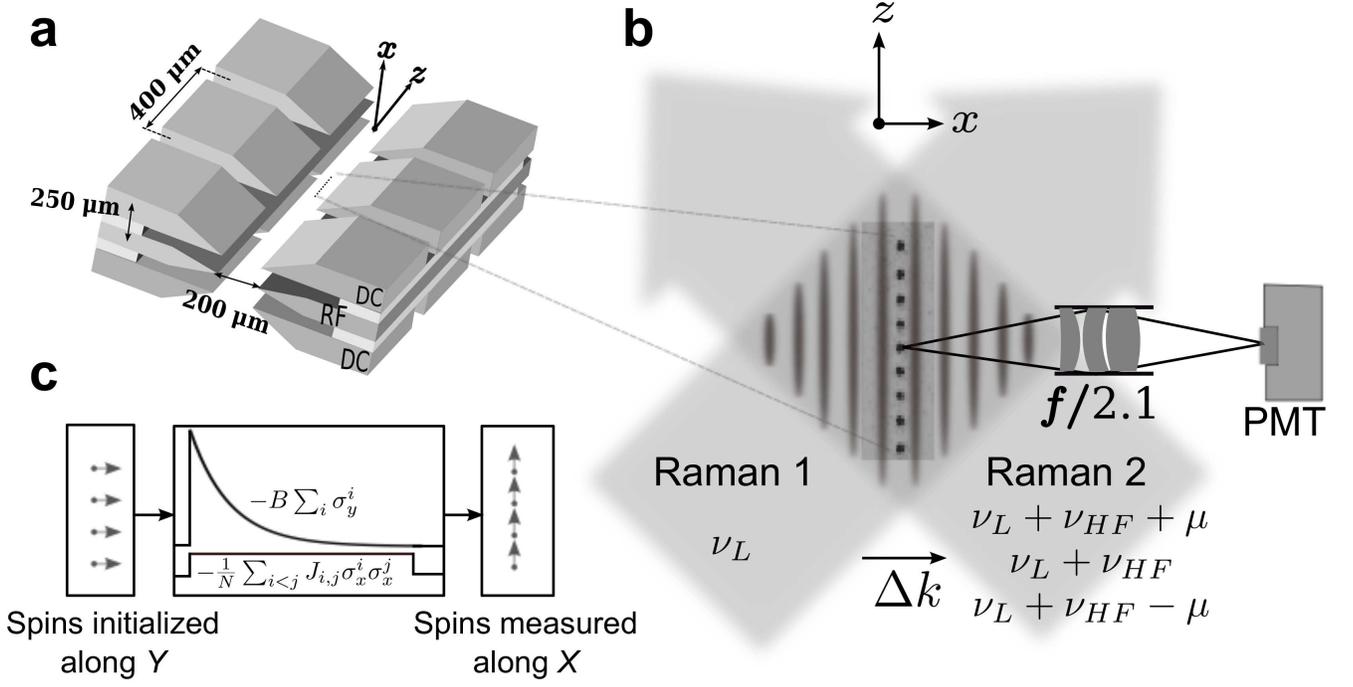}
\end{flushleft}
\caption{\label{fig:schematic}
{\bf{Experimental schematics}} : {\bf{a}}. Schematic of the three-layer linear rf (Paul) trap, with top and bottom layers carrying static potentials and the middle one carrying rf.  {\bf{b}}. Two Raman beams globally address the $^{171}\rm{Yb}^{+}$ ion chain, with their wavevector difference ($\vec{\Delta k}$) along the transverse $(x)$ direction of motion, generating the Ising couplings through a spin-dependent force. The same beams generate an effective transverse magnetic field by driving resonant hyperfine transitions. A CCD image showing a string of nine ions (not in present experimental condition) is superimposed. A photomultiplier tube (PMT) is used to detect spin-dependent fluorescence from the ion crystal.  {\bf{c}}. Outline of quantum simulation protocol. The spins are initially prepared in the ground state of $-B\sum_i\sigma^i_y$, then the Hamiltonian \ref{Ham} is turned on with starting field $B_{0}\gg |J|$ followed by an adiabatic exponential ramping to the final value $B$, keeping the Ising couplings fixed.  Finally the $X-$ component of the spins are detected.}
\end{figure*}

We represent each spin-$\frac{1}{2}$ system by the hyperfine clock states $^{2}S_{1/2}|F=0,m_{F}=0\rangle$ and $|F=1,m_{F}=0\rangle$ of $^{171}\rm{Yb}^{+}$ separated by $\nu_{HF}=$ 12.642821 GHz, which are denoted by 
the eigenstates $\downarrow_{z}$ and $\uparrow_{z}$ of $\sigma_z$, respectively. 
These states are detected by standard spin-dependent resonant fluorescence on the cycling $^{2}S_{1/2}$ to $^{2}P_{1/2}$ transition at 369.5 nm using a photomultiplier tube \cite{OlmschenkYb}.
The ions are trapped along the $z-$axis of a three layer linear Paul trap (see Fig.~\ref{fig:schematic}a) with center of mass (CM) vibrational frequencies of $\nu_x = 4.748$, $\nu_y = 4.641$, and $\nu_z = 1.002$ MHz  along the $x,y$ (transverse) and $z$ (axial) directions, respectively \cite{DesLauriers04}. The modes of motion along $x$ are cooled to near their vibrational ground states and within the Lamb-Dicke regime. 
Off-resonant laser beams address the ions globally, driving stimulated Raman transitions between the spin states and also imparting spin-dependent forces exclusively in the $x$-direction, as depicted in Fig.~\ref{fig:schematic}b \cite{KimPRL, KimNature, EdwardsPRB}.  The Raman beams contain a ``carrier" beat-note at frequency $\nu_{HF}$, which provides an effective uniform transverse magnetic field $B$.  Raman beatnotes at frequencies $\nu_{HF} \pm \mu$ are near motional sidebands and generate a spin-dependent force.  The rf phase difference between the carrier beatnote and the average beatnote of the sidebands is set to $\pi/2$ so that the magnetic field is transverse to the Ising couplings (Eq 1) \cite{KimNature, EdwardsPRB}.  We suppress direct sideband (phonon) excitation by ensuring that the beatnote detuning $\mu$ is sufficiently far from any normal mode frequency \cite{KimPRL}.  This requires that $|\mu-\nu_m|>>\eta_{i,m} \Omega_i$, where $\eta_{i,m}$ is the Lamb-Dicke parameter of the $i$th ion and $m$th normal mode at frequency $\nu_m$ (with $\nu_1=\nu_x$ denoting the CM mode), and $\Omega_i = g_{i}^2/\Delta$ is the carrier Rabi frequency on the $i$th ion.  Here $g_{i}$ is the single photon Rabi frequency of $i^{th}$ ion and $\Delta$ is the detuning of the Raman beams from the $^{2}S_{1/2}-^{2}P_{1/2}$ transition. 

\begin{figure*}[t]
\begin{center}
\includegraphics[width=0.95\linewidth]{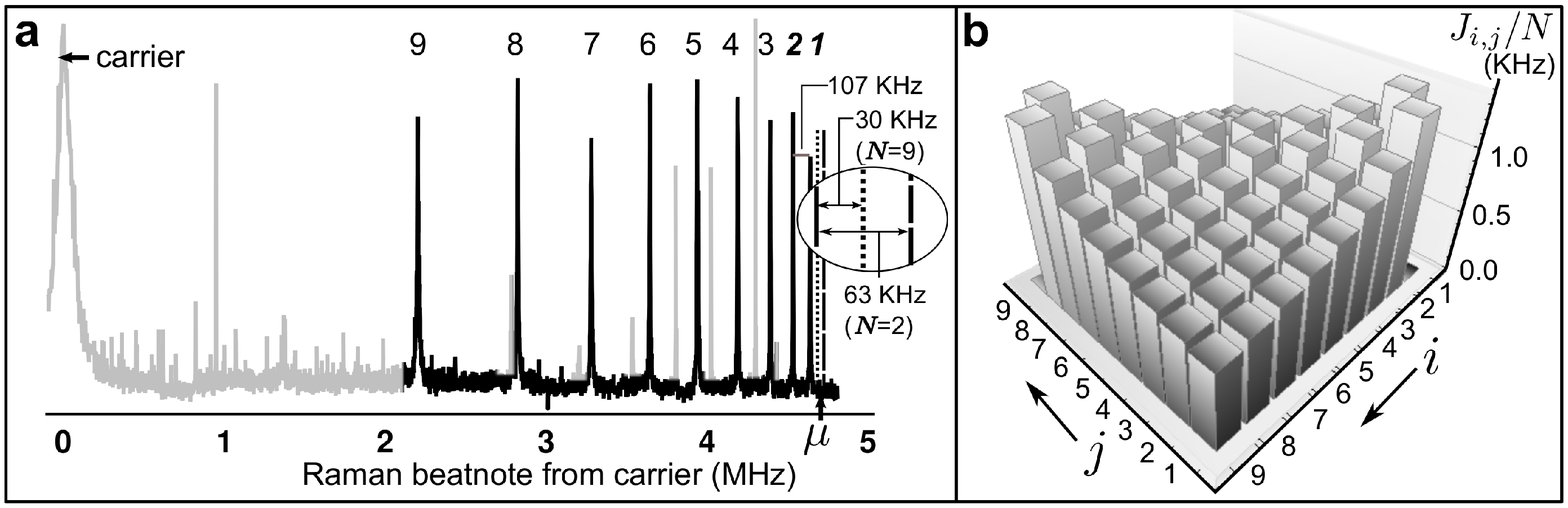}
\end{center}
\caption{\label{fig:Jij}
{\bf{Motional modes and Ising couplings}}: Transverse \cite{ZhuTransverse} vibrational modes  are used in the experiment  to generate Ising couplings according to Eq. \ref{Jij}. {\bf{a}}. Raman sideband spectrum of vibrational normal modes along transverse $x-$direction for nine ions, labeled by their index $m$.  The two highest frequency modes at $\nu_1$ (CM mode) and $\nu_2=\sqrt{\nu_1^2 - \nu_z^2}$ (``tilt" mode) occur at the same position independent of the number of ions.  The dotted and the dashed lines show beatnote detunings of $\mu \approx \nu_1 + 30$ KHz and $\mu \approx \nu_1 + 63$ KHz used in the experiment for $N=9$ and $N=2$ ions respectively. Carrier transition, weak excitation of transverse$-y$ and axial$-z$ normal modes and higher order modes are faded (light grey) for clarity. {\bf{b}}.  Theoretical Ising coupling pattern (Eq. \ref{Jij}) for $N=9$ ions and uniform Raman beams.  The main contribution follows from the uniform CM mode, with inhomogeneities given by excitation through the other nearby modes (particularly the tilt mode). Here, $J_{1,1+r}\sim 1/r^{0.35}$ ($r\ge 1$), as found out empirically. For larger detunings, the range of the interaction falls off even faster with distance, approaching the limit $J_{i,j} \sim  1/|i-j|^3$ for $\mu \gg \nu_1$ \cite{Porras04}.}

\end{figure*}

This results in an Ising interaction between the spins with control parameter $\mu$ that dictates the form of the coupling matrix \cite{KimPRL}
\begin{equation}
J_{i,j} = N\Omega_i\Omega_j \sum_{m=1}^{N}\frac{\eta_{i,m}\eta_{j,m}\nu_m}{\mu^2 - \nu_m^2}.
\label{Jij}
\end{equation}
In the experiment $\Delta\sim 2.7$ THz, $\Omega_i\sim 370$ KHz and we expect $J_{i,j}/N\sim 1$ KHz for the beatnote detuning $\mu$ such that $\mu-\nu_1 \approx 4\eta_{i,1}\Omega_i$, as shown in Fig.~\ref{fig:Jij}a .  This beatnote corresponds to $63$ KHz blue of the CM mode frequency for $2$ ions and $30$ KHz for $9$ ions, as the Lamb-Dicke parameter $\eta_{i,m} \sim \frac{1}{\sqrt{N}}$.  This maintains roughly the same level of  virtual phonon excitation for any number of ions.  The expected Ising coupling pattern for a uniformly illuminated ion chain is shown in Fig.~\ref{fig:Jij}b for $N=9$ ions and the couplings are dominated by uniform contribution of the CM mode. The non-uniformity in the Ising couplings arises from other vibrational modes and a $<1\%$ variation in $\Omega_i$ across the ion chain for $\sim 100$ $ \mu$m $\times 5 \mu$m wide gaussian Raman beams used in the experiment. 

\begin{figure*}
\begin{center}
\includegraphics[width=0.95\linewidth]{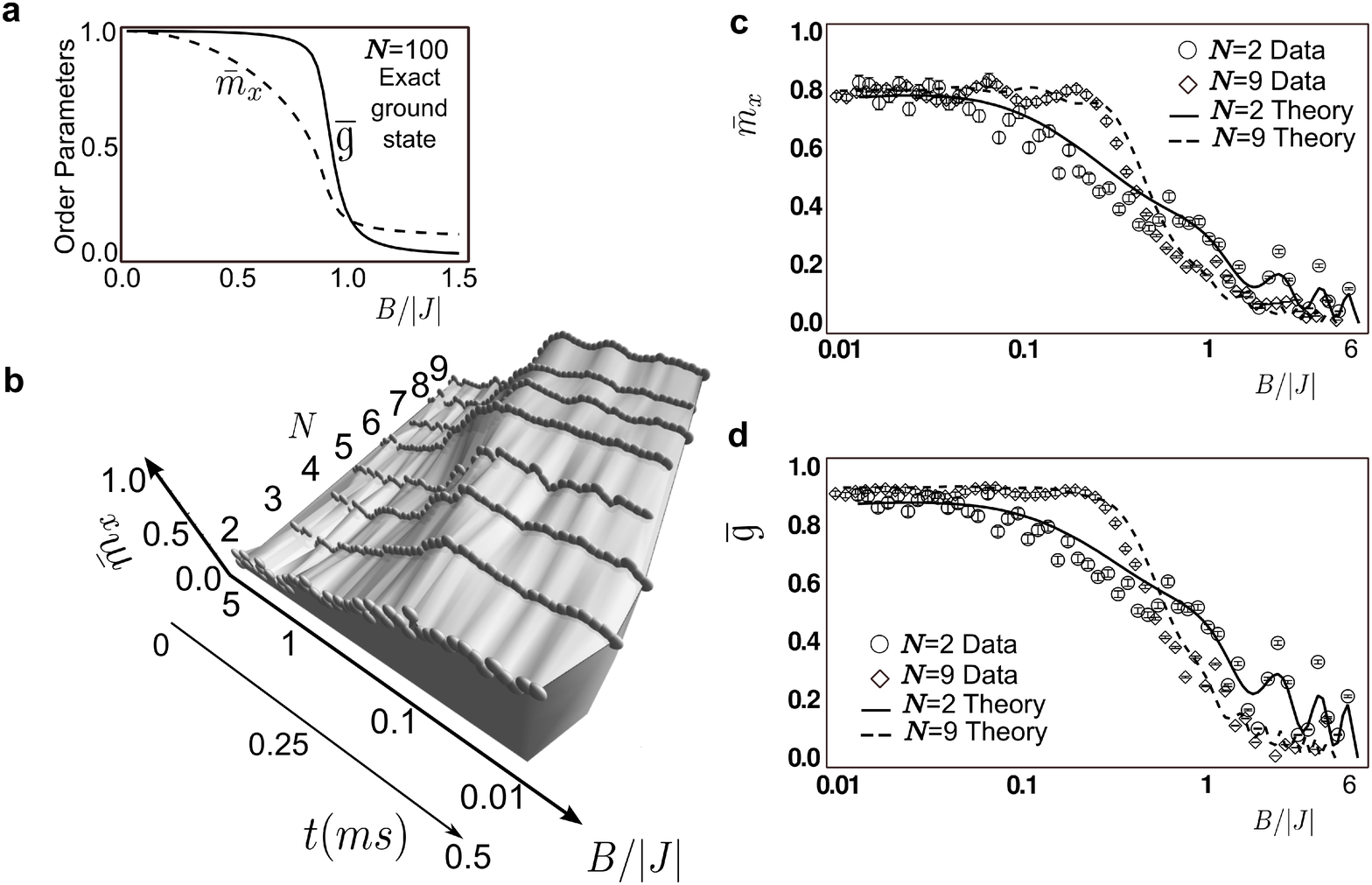}
\end{center}
\caption{\label{fig:QSim_Data}
{\bf{Experimental results of adiabatic quantum simulation}} : 
  {\bf{a}}.  Theoretical values of order parameters  ${\bar{m}}_x$ and $\bar{g}$ are plotted vs $B/|J|$ for a moderately large system ($N=100$ spins) with uniform Ising couplings, in the case of a perfectly adiabatic time evolution. The effective ground state manifold reduces to $N+1$ dimensions in the total spin basis when the Ising couplings are uniform. The scaled Binder cumulant $\bar{g}$ approaches a step function near the transition point $B/|J|=1$ unlike the scaled magnetization ${\bar{m}}_x$, making it experimentally suitable to probe the transition point for relatively small systems. {\bf{b}}. Scaled average absolute  magnetization per site, ${\bar{m}}_x$  vs $B/|J|$ (and simulation time) is plotted for $N=2$ to $N=9$ spins.  As $B/|J|$ is lowered, the spin ordering undergoes a crossover from a paramagnetic to ferromagnetic phase. The crossover curves sharpen as the system size is increased from $N=2$ to $N=9$, anticipating a QPT in the limit of infinite system size. The oscillations in the data arise due to imperfect initial state preparation and non-adiabaticity due to finite ramping time. The (unphysical) $3$D background is shown to guide eyes. {\bf{c}}.  Magnetization data for $N=2$ spins (circles) is contrasted with $N=9$ spins (diamonds), near $B/|J|=1$.  The data deviate from unity at  $B/|J|=0$ by $\sim 20\%$, predominantly due to decoherence from spontaneous emission in Raman transitions and additional dephasing from Raman beam intensity fluctuation, as discussed in the text.  The theoretical time evolution curves (solid line for $N=2$ and dashed line for $N=9$ spins) are calculated by averaging over 10,000 quantum trajectories, as described in the Methods section.  {\bf{d}}. Scaled Binder cumulant ($\bar{g}$) data (circles and diamonds) and time evolution theory curves (solid and dashed lines) are plotted for $N=2$ and $N=9$ spins. At $B/|J|=0$ the data deviate by $\sim 10\%$ from unity, due to decoherence as mentioned before. 
}

\end{figure*}

The experiment proceeds as follows (Fig.~\ref{fig:schematic}c). We initialize the spins to be aligned to the $Y-$ direction of the Bloch sphere by optically pumping to $|\downarrow_z\downarrow_z...\downarrow_z\rangle$ and then coherently rotating the spins through $\pi/2$ about the Bloch $X-$axis with a carrier Raman transition.  Next we switch on the  Hamiltonian $H$ with effective magnetic field $B_0\sim 5|J|$ so that the spins are prepared predominantly in the ground state. Then we exponentially ramp down the effective magnetic field with a time constant of 80 $\mu s$ to a final value $B$, keeping the Ising couplings fixed.  We finally measure the spins along the Ising $(X)$ direction by coherently rotating the spins through $\pi/2$ about the Bloch $Y-$axis before fluorescence detection.
We repeat the experiment $\sim 1000 N$ times and generate a histogram of fluorescence counts and fit to a weighted sum of basis functions to obtain the probability distribution $P(s)$ of the number of spins in state ($|\uparrow\rangle$), where $s=0,1,...N$, as described in the methods section.

We can generate several magnetic order parameters of interest from the distribution $P(s)$, showing transitions between different spin orders.  
One order parameter is the average absolute 
\footnote{ Hamiltonian \ref{Ham} has a global time reversal symmetry of $\{\sigma^i_x\to-\sigma^i_x$, $\sigma^i_z\to-\sigma^i_z$, $\sigma^i_y\to\sigma^i_y\}$ and this does not spontaneously break for a finite system, necessitating the use of average $\it{absolute}$ value of the magnetization per site along the Ising direction as the relevant order parameter. } 
magnetization per site along the Ising direction, $m_{x}=\frac{1}{N}\sum\limits_{s=0}^{N}|N-2s|P(s)$.  For a large system, this parameter shows a second order phase transition, or a discontinuity in its derivative with respect to $B/|J|$.  On the other hand, the fourth-order moment of the magnetization or Binder cumulant (BC)  $g=\sum\limits_{s=0}^{N}(N-2s)^{4}P(s)/\left(\sum\limits_{s=0}^{N}(N-2s)^{2}P(s)\right)^{2}$  \cite{BinderPhysikB,Binder81} becomes a step function at the QPT and should therefore be more sensitive to the phase transition.  We illustrate this point by plotting the exact ground state order in the simple case of uniform Ising couplings for a moderately large system ($N=100$) in Fig. \ref{fig:QSim_Data}a. In figure \ref{fig:QSim_Data}b-d we present data  for these two order parameters as $B/|J|$ is varied in the adiabatic quantum simulation, with the order parameters properly scaled to account for trivial finite size effects, as described in the methods section.  Fig.~\ref{fig:QSim_Data}b shows scaled magnetization, ${\bar{m}}_x$ for $N=2$ to $N=9$ spins that depict the sharpening of the crossover curves from paramagnetic to ferromagnetic spin order with increasing system size.  The linear time scale indicates the exponential ramping profile of the (logarithmic) $B/|J|$ scale. Fig.~\ref{fig:QSim_Data}c-d compares the two extreme system sizes in the experiment, $N=2$ and $N=9$ and clearly shows the increased steepness for larger system size. The scaled magnetization ${\bar{m}}_x$ is suppressed by $\sim 25\%$ (Fig.~\ref{fig:QSim_Data}b,c) and the scaled BC $\bar{g}$ is suppressed by $\sim 10\%$ (Fig.~\ref{fig:QSim_Data}d) from unity at $B/|J|=0$, predominantly due to decoherence from off-resonant spontaneous emission and additional dephasing due to intensity fluctuations in Raman beams during the simulation. 

We compare the data shown in Fig.~\ref{fig:QSim_Data}c-d to the theoretical evolution taking into account experimental imperfections and errors discussed below, including spontaneous emission to the spin states and states outside the Hilbert space, and additional decoherence.  The evolution is calculated by averaging $10,000$ quantum trajectories. This takes only one minute on a single computing node for $N=2$ spins and approximately 7 hours, on a single node, for $N=9$ spins. Extrapolating from this calculation suggests that averaging $10,000$  trajectories for $N=15$ spins would require 24 hours on a 40 node cluster,  indicating the inefficiency of classical computers to simulate even a small quantum system.

\begin{figure*}
 \includegraphics[width=0.95\linewidth]{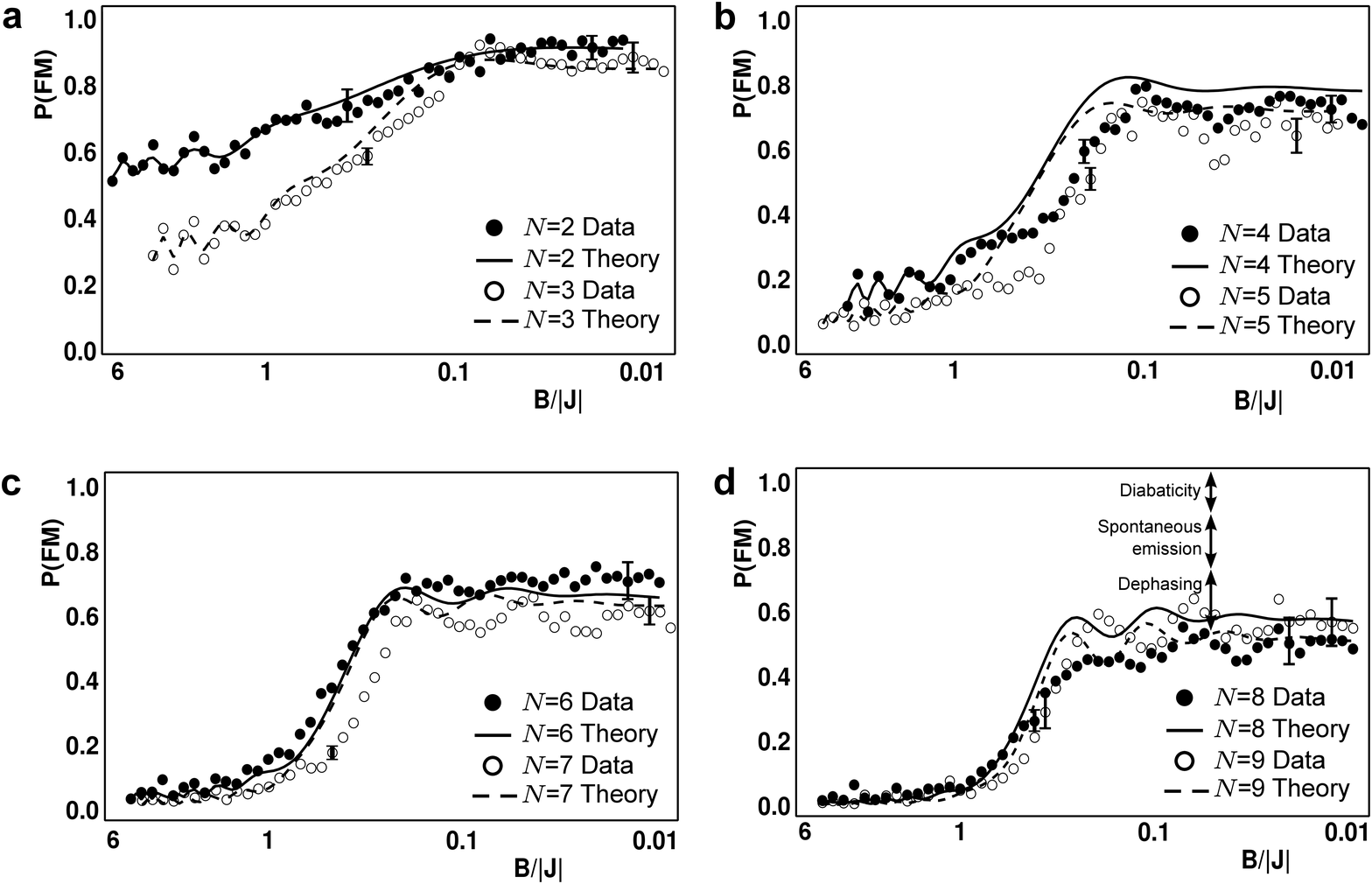}

 \caption{\label{fig:PFM_all}
{\bf{Suppression of P(FM) with increasing number of spins}} : {\bf{a-d}.} Ferromagnetic order P(FM)=P(0)+P(N) is plotted vs $B/|J|$ for $N=2$ to $N=9$ spins. The circles are experimental data and the lines are theoretical results including decoherence and imperfect initialization. As this quantity includes only two of $2^{N}$ basis states random spin-flips and other errors degrade it much faster than the magnetization and Binder cumulant. The representative detection error bars are shown on a few points for each $N$. The P(FM) reduces from $\sim0.9$ to $\sim0.55$ as the system size is increased from two to nine. The principle contribution to this degradation is decoherence, predominantly due to spontaneous emission from intermediate $^{2}P_{1/2}$ states in the Raman transition and additional dephasing primarily due to intensity fluctuations in Raman beams. Shown in {\bf{d}} is an estimated breakdown of the suppression of P(FM) from various effects for $N=9$ spins. Non-adiabaticity due to finite ramping speed, spontaneous emission and additional dephasing due to fluctuating Raman beams suppress P(FM) by $\sim8\%$, $\sim 18\%$ and $\sim24\%$ respectively from unity ($B/|J|\to0$).}

\end{figure*}

A faithful quantum simulation requires an excellent understanding of errors and their scaling, especially when the underlying problem is otherwise intractable.  We characterize errors in the current simulation by 
plotting the observed parameter P(FM)=P(0)+P(N) for $N=2$ to $N=9$ spins in Fig.~\ref{fig:PFM_all} and comparing with theory. 
The parameter $P(FM)$ involves only two of the $2^{N}$ basis states and is therefore more sensitive to errors compared with the order parameters ${\bar{m}}_x$ and $\bar{g}$. For instance, at $B/|J|=0$ in Fig.~\ref{fig:QSim_Data}b-d and Fig.~\ref{fig:PFM_all}a-d we find that  ${\bar{m}}_x$ and  $\bar{g}$ do not change appreciably with system size, but P(FM) degrades to $\sim 0.55$ for $N=9$ spins from $\sim 0.9$ for $N=2$.

There are several primary sources of experimental error.  Diabaticity due to finite ramping speed and error in initialization is estimated to suppress P(FM) by $\sim 3\%$  for $N=2$ to $\sim 8\%$ for $N=9$. This also gives rise to oscillations seen in the data (Fig.~\ref{fig:QSim_Data}b-d and Fig.~\ref{fig:PFM_all}a-d). A major source of error is the spontaneous emission from Raman beams which amounts to a  $\sim 10\%$ spontaneous emission probability per spin in $1$ ms. Spontaneous emission dephases and randomizes the spin state and loosely behaves like a `spin temperature' in this system, though the spins do not fully equilibrate with the `bath' and the total probability of spontaneous emission increases linearly during the quantum simulation. 
In addition, each spontaneous emission event populates other states outside of the Hilbert space of each spin with a probability of $1/3$.
Spontaneous emission errors grow with increasing system size, which also suppresses P(FM) order with increasing $N$, as seen  in Fig.~\ref{fig:PFM_all}a-d. We theoretically estimate the suppression of P(FM) due to diabaticity and spontaneous emission together by averaging over quantum trajectories to be $\sim 7\%$ for $N=2$ spins  and $\sim 26\%$ for $N=9$ spins. Intensity fluctuations on the Raman beams during the simulation modulate the ac Stark shift on the spins and dephase the spin states, which causes additional diabaticity and degrades the final ferromagnetic order. When we introduce a theoretical dephasing rate of $0.3$ per ms per ion in the quantum trajectory computation the predicted suppression of P(FM) increases to $\sim 9\%$ for $N=2$ and $\sim 50\%$ for $N=9$. 

Imperfect spin detection efficiency contributes $\sim5-10\%$ error in P(FM).   Fluorescence histograms for P(0) and P(1) have a $\sim 1\%$ overlap (in detection time of 0.8 ms) due to off-resonant coupling of the spin states to the $^{2}P_{1/2}$ level. This prevents us from increasing detection beam power or photon collection time to separate the histograms. Detection error in the data include uncertainty in fitting the observed fluorescence histograms to determine $P(s)$, intensity fluctuations and finite width of the detection beam.

The role of phonons in the results of quantum simulation is investigated both experimentally and numerically. We perform another set of experiments with $\mu-\nu_{1}=63$ KHz for all $N$, which  amounts to reducing the phonon excitation as the number of spins $N$ is increased, since the Lamb-Dicke parameter  $\eta_{i,m} \sim \frac{1}{\sqrt{N}}$. We do not note any appreciable difference (beyond the margin of experimental errors) in spin population with the results reported here. In the presence of the applied effective magnetic field, phonon modes are coherently populated and generally exhibit spin-motional entanglement.  However, these phonons do not alter the spin ordering and hence preserve spin-spin correlation, even if the entanglement between spin states is partly destroyed when measurement traces over phonon states.  Thus a standard GHZ type witness operator will not show entanglement beyonds a few spins. 

In this experiment we have qualitatively observed sharpening of crossover curves which indicates the onset of a quantum phase transition as the system size increases. This scheme can be scaled up to larger number of spins where it is  possible to quantitatively estimate finite size effects e.g., scaling in critical exponents near the phase transition point\cite{Fisher72}. The primary challenges in experimenting with larger system sizes include the spontaneous emission as described above, the requirement of larger optical power to maintain same level of Ising couplings,  and non-adiabatic effects due to shrinking gap between ground and first excited state  \cite{Dusuel_Ising_Gap,Caneva08}. 
One solution is to implement a high power laser with a  detuning far from the $^{2}P$ energy levels, which would minimize spontaneous emission while maintaining the same level of Ising couplings. This would also allow versatility in varying the Ising interaction (together with the effective external field) during the simulation, as the differential A.C. stark shift between spin states is negligible for a sufficiently large detuning. The coherence time increases in the absence of spontaneous emission, allowing for a longer simulation time necessary to preserve adiabaticity as the system grows in size. Recently Raman transitions have been driven using a mode-locked high power pulsed laser at a wavelength of 355 ~nm, which is optimum for $^{171}\rm{Yb}^+$ wherein the ratio of differential AC Stark shift to Rabi frequency is minimized and spontaneous emission probabilities per Rabi cycle are $<10^{-5}$ per spin \cite{CampbellPRL10}.

With this system, it is possible to engineer different Ising coupling patterns by controlling the Raman beatnote detuning $\mu$ and observe interesting spin ordering and phase transitions some of which can be very sharp, or first order \cite{LinKink}. Long range interactions in this spatially one dimensional system allow for simulating multi-dimensional spin models by selectively exciting vibrational modes using multiple Raman beatnote detunings. With additional laser beams this scheme can potentially simulate more complicated and higher spin-dimensional Hamiltonians like $XY$ and $XYZ$ models, which map onto nontrivial quantum Hamiltonians such as the Bose-Hubbard Hamiltonian.

We thank D. Huse for help with theoretical understanding of the quantum Ising model and finite size scaling. This work is supported under Army Research Office (ARO) under Award No. 911NF0710576 with funds from the DARPA Optical Lattice Emulator (OLE) Program, IARPA under ARO contract, the NSF Physics at the Information Frontier (PIF) Program, the European Program on Atomic Quantum Technologies (AQUTE), and the NSF Physics Frontier Center at JQI.

\bibliography{QSim_TFIM_arXiv}

\section*{Methods}
 
\subsection{Detection of spin states.}
The spin states are detected by spin-dependent fluorescence signals collected through $f/2.1$ optics by a photomultiplier tube. Spin state $|\uparrow_z\rangle$ is resonantly excited by the 369.5 nm detection beam and fluoresces from $^{2}P_{1/2}$ states, emitting Poisson distributed photons with mean $\sim 12$ in 0.8 ms. This state appears as `bright' to PMT. The detection light is  far off-resonant to spin state $|\downarrow_z\rangle$ and this state appears `dark' to the PMT.  However, due to weak off-resonant excitation bright state leaks onto dark state, altering the photon distribution \cite{ActonThesis}. Unwanted scattered light from optics and trap electrodes also alter the photon distribution. We construct the basis function for $s$ bright ions by convolution techniques, and include a $5\%$ fluctuation in the intensity of detection beam, which is representative of our typical experimental conditions. We then fit the experimental data to these basis functions, and obtain probabilities $P(s)$. Mean photon counts for dark ($m_{D}$) and bright states ($m_B$) are used as fitting parameters so as to minimize  the error residues. 

The error bars to the order parameters are estimated conservatively in the following way. Say, best fitting is obtained for $m_D={\bar{m}}_D$ and $m_B={\bar{m}}_B$. The fitting error in those quantities are $\delta m_D$ and $\delta m_B$ respectively. We generate histograms of the order parameters by extracting $P(s)$  $\sim400$ times with mean dark and bright state counts chosen randomly (Monte Carlo method of error analysis) from a Gaussian distribution with mean ${\bar{m}}_D$, ${\bar{m}}_B$ and standard deviations $\delta m_D$,  $\delta m_B$ respectively. The histogram is fitted to a Gaussian and the width is chosen to represent the random error due to fit. The uncertainty in amount of fluctuation of the detection beam power during the experiment is  conservatively included in the error analysis by repeating the fitting process for a range of fluctuations.  The finite width of the detection beam is taken care of by modeling the Gaussian beam having a three step intensity profile with appropriate intensity ratios. 

\subsection{Scaled Order Parameters.}
To characterize the spin orders we use different order parameters in the experiment, namely the average absolute magnetization per site ($m_x$) and the Binder cumulant ($g$). When the spins are polarized along the $Y-$direction of the Bloch sphere, the distribution of total spin along $X-$direction is Binomial and approaches a Gaussian (with zero mean) in the limit of $N\to\infty$.  For system size of $N$, $m_{x}$ takes on theoretical value of $m^{0}_{x,N}=\frac{1}{N2^{N}}\sum\limits_{n=0}^{N} {^{N}}C_{n} |N-2 n|$ in the perfect paramagnetic phase ($B/|J|\to\infty$) and unity in the other limit of $B/|J|=0$. In Fig.~\ref{fig:QSim_Data}b and \ref{fig:QSim_Data}c we rescale $m_{x}$ to ${\bar{m}}_{x}=\left(m^{0}_{x,N}-m_{x}\right)/\left(m^{0}_{x,N}-1\right)$ which should ideally be zero in perfect paramagnetic phase and unity in perfect ferromagnetic phase for any $N$.  This accounts for the `trivial' finite size effect due to the difference between Binomial and Gaussian distribution. Similarly the Binder Cumulant $g$ is scaled to $\bar{g}=\left(g^{0}_{N}-g\right)/\left(g^{0}_{N}-1\right)$ in Fig.~\ref{fig:QSim_Data}d, where $g^{0}_N=3-2/N$ is the theoretical value of $g$ for $B/|J|\to\infty$. 

\subsection{Quantum Monte-Carlo Simulations.}
Quantum trajectories are generated by numerically integrating the Schr\"odinger equation, with Hamiltonian (1), while simultaneously executing quantum jumps to account for spontaneous emission and decoherence. Spontaneous emission from ion $i$ either localizes the spin of the ion, projecting it into $2S_{1/2}|F=0,m_F=0\rangle$ (spin state $|\downarrow_z\rangle$) or $2S_{1/2}|F=1,m_F=0\rangle$ (spin state $|\uparrow_z\rangle$), or it projects the ion into $2S_{1/2}|F=1,m_F=1\rangle$, in which case ion $i$ is factored out of the Schr\"odinger evolution, though it is counted as spin up at the time of measurement. Decoherence is modeled by the quantum jump operator $\sigma_x$; thus a jump for ion $i$, $|\psi\rangle\to\sigma_x^i|\psi\rangle$, introduces a $\pi$ phase shift between the spin states $|\uparrow\rangle$ and $|\downarrow\rangle$ (in $X-$basis) . Jump rates are taken to be fixed and equal for all ions. To determine the entangled state of the spin ensemble after a spontaneous emission, e.g.~from ion $i$, we assume that the ground state configuration prior to emission,
\begin{eqnarray*}
 |\uparrow_z\rangle_i\prod_{j\neq i}\left(\alpha_{i,j}|\uparrow_z\rangle_j+\beta_{i,j}|\downarrow_z\rangle_j\right)+\\
|\downarrow_z\rangle_i\prod_{j\neq i}\left(\gamma_{i,j}|\uparrow_z\rangle_j+\delta_{i,j}|\downarrow_z\rangle_j\right)
\end{eqnarray*}

is mapped, by the far detuned Raman beams, into a very small excited-state contribution to the overall system entangled state,
$$
\lambda|2P_{1/2}\rangle_i\prod_{j\neq i}\left[(\alpha_{i,j}+\gamma_{i,j})|\uparrow_z\rangle_j+(\beta_{i,j}+\delta_{i,j})|\downarrow_z\rangle_j\right)],
$$
with $\lambda\ll1$ proportional to the amplitude of the Raman beams and inversely proportional to their detuning. The (unnormalized) state after the emission is
$$
|{\rm ?}\rangle_i\prod_{j\neq i}\left[(\alpha_{i,j}+\gamma_{i,j})|\uparrow_z\rangle_j+(\beta_{i,j}+\delta_{i,j})|\downarrow_z\rangle_j\right)],
$$
where $|{\rm ?}\rangle_i$ is $|\uparrow_z\rangle_i$, $|\downarrow_z\rangle_i$, or the factored state $2S_{1/2}|F=1,m_F=1\rangle$.

\end{document}